\documentclass[showpacs,prl,twocolumn,
superscriptaddress]{revtex4}
\usepackage{graphicx}

\newcommand{\be}{\begin{equation}}
\newcommand{\ee}{\end{equation}}
\newcommand{\ba}{\begin{eqnarray}}
\newcommand{\ea}{\end{eqnarray}}
\newcommand{\bd}{\begin{description}}
\newcommand{\ed}{\end{description}}

\renewcommand{\iota}{{\bf 1}}
\def\rellow#1#2{Mathrel{Mathop{\kern 0pt #1}\limits_{#2}}}

\begin{document}

\preprint{USACH-TH-2005-XXX }
\title{Scattering of Solitary Waves in Granular Media}
\author{Lautaro Vergara}
\email{lvergara@lauca.usach.cl}
\affiliation{Departamento de F\'{\i}sica, Universidad de Santiago de Chile, Casilla 307,
Santiago 2, Chile }
\date{today}

\begin{abstract}
A detailed numerical study of the scattering of solitary waves by a barrier,
in a granular media with Hertzian contact, shows the existence of secondary
multipulse structures generated at the interface of two "sonic vacua", which
have a similar structure as the one previously found by Nesterenko and
coworkers.
\end{abstract}

\maketitle

\section{Introduction}

Nesterenko \cite{Nesterenko} noticed that the propagation of a perturbation
in a chain of beads in Hertzian contact possesses soliton-like features.
Several studies, theoretical as well as experimental, \cite{Wattis}, \cite%
{Sen1}, \cite{Coste}, \cite{Hinch} have confirmed the existence of
such soliton-like pulses. Despite the great deal of recent work on
the subject
\cite{Sen1,Hinch,Manciu,SenMan,Coste,Lee,Hong,SenProc,Naka,Rosas,Sen2,Nest},
the physics of granular media remains a challenge and new effects
are there to be discovered and studied.

The simplest granular systems are one-dimensional chain of elastic spheres.
It is known \cite{Nesterenko} that in the regime where Hertz's static
solution for the contact of elastic spheres applies, the spheres may be
considered as point masses interacting through massless nonlinear springs
with elastic force $F=k\delta ^{3/2}$, where $\delta $ is the overlap of
contacts and $k$ is the spring constant (a function of the material
properties). Let $v_{i}(t)$ represents the displacement of the center of the
$i$-th sphere from its initial equilibrium position, and assume that the $i$%
-th sphere, of mass $m$, has neighbor of different radius (and/or mechanical
properties). Then, in absence of load and in a frictionless medium, the
equation of motion for the $i$-th sphere reads
\begin{equation}
m\frac{d^{2}v_{i}}{dt^{2}}%
=k_{1}(v_{i-1}-v_{i})^{3/2}-k_{2}(v_{i}-v_{i+1})^{3/2},  \label{uno}
\end{equation}%
where it is understood that the brackets take the argument value if they are
positive and zero otherwise, ensuring that the spheres interact only when in
contact.

Nesterenko, Lazaridi and Sibiryakov \cite{Nesterenko2} studied
experimentally as well numerically the interaction of a solitary wave with
the boundary of two "sonic vacua" using a setup similar as the one shown in
Figure 1, with a wall at the right hand side. They also analyzed the case
where the small and big beads are interchanged.

In this work we make a detailed numerical study of the propagation of
solitary waves in a linear chain of beads composed of two types of bead
sizes, as shown in Figure 1, repeating the analysis done in \cite%
{Nesterenko2}. It will be assumed that all spheres are made of the same
material and that both ends of the chain are free to move.

\section{Numerical study and results}

Consider a set of spheres with two different radius $a$ and $b$. It is known
that adjacent spheres of radius $a$ and $b$ will interact with a force $%
F=k\delta^{3/2}$, where
\begin{equation}
k=\frac{\sqrt{ab/(a+b)}}{2\theta },  \label{dos}
\end{equation}
with
\begin{equation}
\theta =\frac{3(1-\nu ^{2})}{4E}  \label{tres}
\end{equation}
and $E$ is the Young modulus and $\nu $ the Poisson ratio of the bead
material.

We will consider the scattering of solitary waves in a setup as that shown
in Fig. 1, similar to the one used in \cite{Nesterenko2}\footnote{%
A series of experiments with similar setups have been done by F.
Melo and coworkers at the Nonlinear Physics Laboratory of the
Physics Department of Universidad de Santiago de Chile}, where
there are $M$ beads, of which $N$ has radius $a$ and mass $m_{1}$
and $L$ radius $b$ and mass $m_{2}$; $a<b$. The displacements of
the beads are governed by a set of eqs. of motion that can be
readily obtained from the successive application of eq.
(\ref{uno}), having in mind that the eq. of motion for the first
(resp. the last) sphere only includes the second (resp. the first)
term, in case there is no wall (as we here assume).

%********|*********|*********|*********|*********|*********|*********|****
%********|*********| Fig1:           |*********|*********|****
%********|*********|*********|*********|*********|*********|*********|****
\begin{figure}[tbp]
\centering %\vspace{-.5cm}
\hspace{-.5 cm} \includegraphics[width=.35\textwidth]{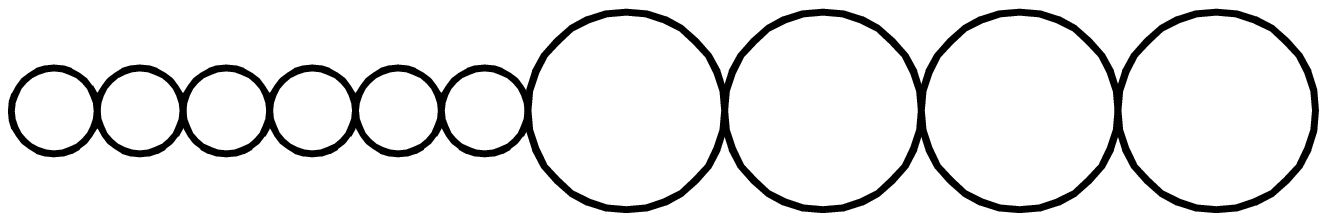} \vspace{%
-.2cm}
\caption{{}}
\label{panel}
\end{figure}
%********|*********|*********|*********|*********|*********|*********|****

For numerical convenience, we rescale the variables, defining
\begin{equation}
u_{i}=\left( \frac{k_{1}}{m_{1}}\right) ^{-2}v_{i}.
\end{equation}

In the first part of our analysis of the effect of the interface
on the dynamics of solitary waves, the parameters of the system
are chosen as $N=60$, $L=60$, $a=1, b=2$, $m_{1}/m_{2}=1/8$. We
assume that initially all beads are at rest, except for the first
bead at the left side of the chain. This bead is supposed to have
a nonzero value of velocity in order to generate the soliton-like
perturbation in the chain. That is, we assume that the initial
conditions are
\begin{eqnarray*}
u_{i}(0) &=&0,\text{ \ }i=1,\ldots ,M,\text{ \ }\dot{u}_{1}(0)=1 \\
\dot{u}_{1}(0) &=&0,\text{ \ }i=2,\ldots ,M,\label{init}
\end{eqnarray*}

In this case, the scattering process is as follows: the solitary
wave arrives at the interface between the two "sonic vacua" and it
splits into a transmitted and a reflected solitary wave as already
observed in \cite{Nesterenko2}. In addition to what reported
there, we have observed the presence of extra multipulse
structures.
%********|*********|*********|*********|*********|*********|*********|****
%********|*********| Fig2:           |*********|*********|****
%********|*********|*********|*********|*********|*********|*********|****
\begin{figure}[h]
%\centering
\vspace{-.5cm} \hspace{-.5 cm}
\includegraphics[width=.35\textwidth]{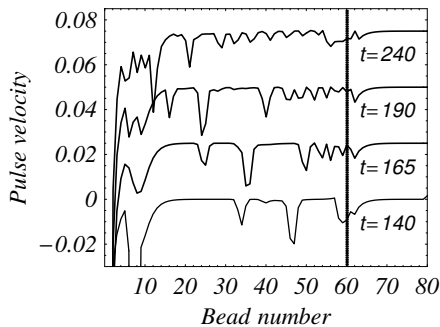} \vspace{-.2cm}
\caption{{Multipulse structure produced by the scattered solitary
wave (by a setup as that shown in Fig. 1) as a function of the
bead number, at different times. The vertical line denotes the
interface position.}} \label{panel1}
\end{figure}

In Figure 2 we have plotted the velocity of the pulses as a
function of the bead number for different times $t=140, 165, 190,
220$ and plots where shifted by $0.025$ for clarity. At $t=140$ it
is observed (part of) the reflected solitary wave and behind it
two pulses appear. In addition, an unresolved perturbation around
the interface is present. As shown in the figure, at later times
multipulse structures appears.

The perturbations in the multipulse structure possess much less
energy than the solitary waves; comparing the velocity of the
reflected solitary wave with the velocity of the first pulse in
the multipulse structure shows that in the case at hand, the
perturbation has around 3\% of the velocity of the solitary wave.

%********|*********|*********|*********|*********|*********|*********|****
%********|*********| Fig3:  |*********|*********|****
%********|*********|*********|*********|*********|*********|*********|****
\begin{figure}[h]
%\centering
\vspace{-.5cm} \hspace{-.5 cm}
\includegraphics[width=.35\textwidth]{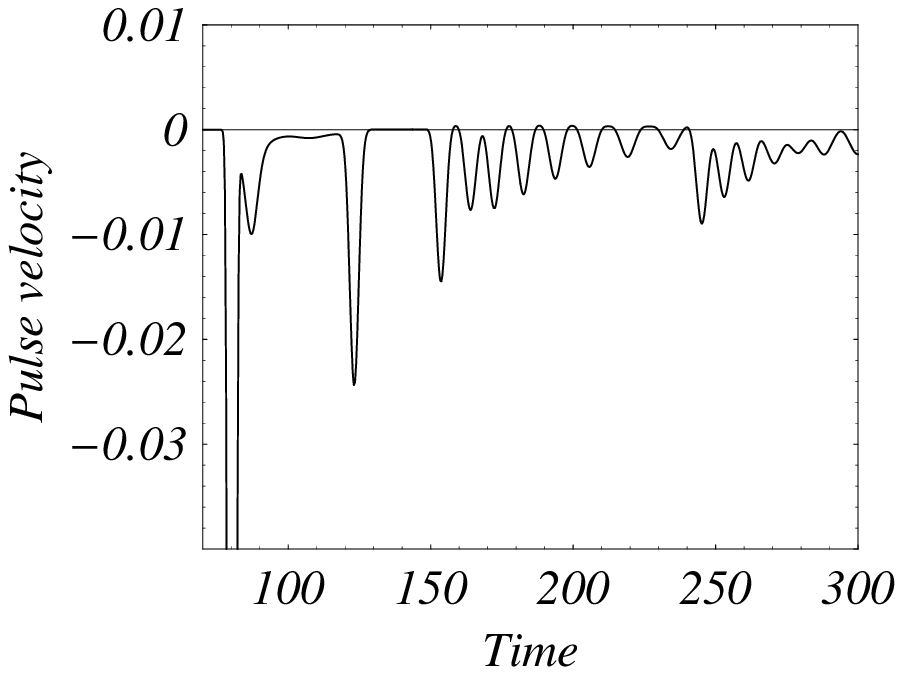} \vspace{-.2cm}
\caption{{Velocity of bead 54 as a function of time.}}
\label{panel1}
\end{figure}

Figure 3 shows the velocity of bead 54 as a function of time. The
large peak corresponds to the reflected solitary wave, while the
velocity of the incident solitary wave is not shown. The two
multipulse structures observed here have they counterpart in the
structures observed at $t=240$ in Figure 2.

It is also interesting to study the dependence of the effects that
we have observed on the strength of the impact by the first bead.
To that end we use as initial conditions
\begin{eqnarray*}
u_{i}(0) &=&0,\text{ \ }i=1,\ldots ,M,\text{ \ }\dot{u}_{1}(0)=3 \\
\dot{u}_{1}(0) &=&0,\text{ \ }i=2,\ldots ,M.
\end{eqnarray*}

%********|*********|*********|*********|*********|*********|*********|****
%********|*********| Fig3:  |*********|*********|****
%********|*********|*********|*********|*********|*********|*********|****
\begin{figure}[h]
%\centering
\vspace{-.5cm} \hspace{-.5 cm}
\includegraphics[width=.35\textwidth]{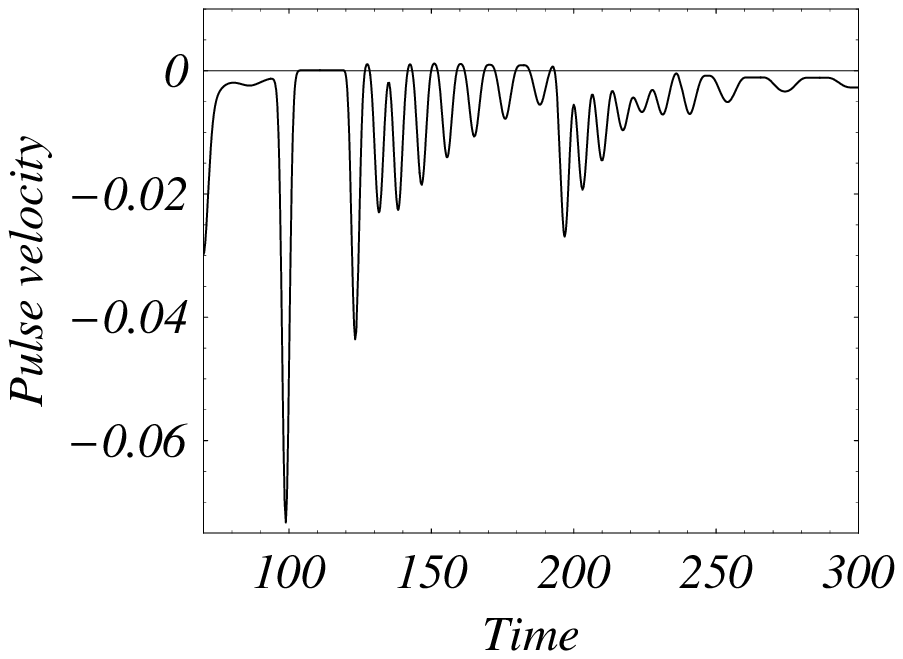} \vspace{-.2cm}
\caption{{Velocity of bead 54 as a function of time.}}
\label{panel1}
\end{figure}

Figure 4 shows the dependence of velocity of beads 54 as a
function of time, for these initial conditions. As expected, all
perturbations are enhanced and because of the larger magnitude of
the initial velocity, perturbations are shifted to earlier times
as compared with Figure 3. No big differences are found between
both cases, indicating that (at least to these orders of
magnitude) the impact velocity has no incidence in the the
appearance of the multipulse structures.

Before continuing with our a analysis we remark that since the
effects that we observe are small it is important to be sure of
the accuracy of the numerics. We have numerically studied the
system of equations by using an explicit Runge-Kutta method of 5th
order based on the Dormand-Prince coefficients, with local
extrapolation. As step size controller we have used the
proportional-integral step control, which gives a smooth step size
sequence. We have tested our numerics by comparing with results
given in the literature as the one shown in \cite{Hinch} and
\cite{Chatterjee}, obtaining a good agreement with them. In
addition, as a crude check of numerical integration accuracy we
have found that the final kinetic energy of the transmitted
solitary wave in the heavy system differs from its initial kinetic
energy in that system by $10^{-7}$.

When studying the dependence of the multipulse structure as a
function of the ratio between the beads, one observes that the
number of pulses in the multipulse structure increases when the
mass ratio $m_{1}/m_{2}$ decreases and viceversa. Also notice that
for lower mass ratios $m_{1}/m_{2}$ the multipulse structure take
more time to appear. This is shown in Figures 5 where the velocity
of
bead 54 as a function of time is plotted. Mass ratios are chosen as $%
m_{1}/m_{2}=1/1.6^{3}$, $m_{1}/m_{2}=1/2^{3}$ and $m_{1}/m_{2}=1/2.5^{3}$.
The curves corresponding to the last two mass ratios have been shifted by $%
0.05$ units for clarity.

%********|*********|*********|*********|*********|*********|*********|****
%********|*********| Fig4:  |*********|*********|****
%********|*********|*********|*********|*********|*********|*********|****
\begin{figure}[h]
\centering
%\vspace{-.5cm} \hspace{-.5 cm}
\includegraphics[width=.35\textwidth]{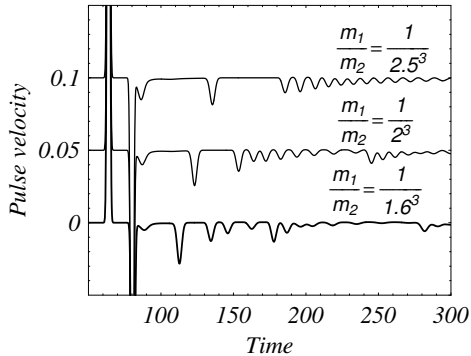}
\vspace{-.2cm} \caption{{Velocity of bead 54 as a function of time
for three different mass ratio.}} \label{panel1}
\end{figure}

One could ask whether similar structures also appear in the case
when the solitary wave moves from the "heavy" system to the
"light" system (from
right to left in Figure 1). This system was first analyzed in \cite%
{Nesterenko2} and here we reanalyze their numerical calculation.
To this end we choose the system's parameters as $a=2, b=1$,
$m_{1}/m_{2}=8$, with 30 big beads and 170 small ones and the
initial conditions (\ref{init}).

%********|*********|*********|*********|*********|*********|*********|****
%********|*********| Fig2:           |*********|*********|****
%********|*********|*********|*********|*********|*********|*********|****
\begin{figure}[h]
%\centering
\vspace{-.5cm} \hspace{-.5 cm}
\includegraphics[width=.35\textwidth]{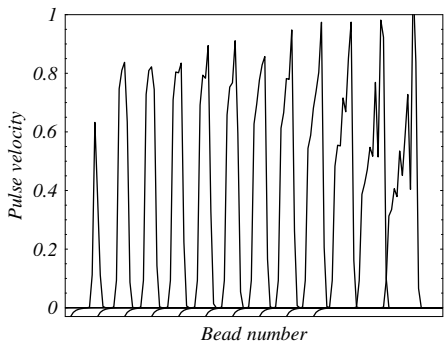} \vspace{-.2cm}
\caption{{Snapshots at times ranging from $t=34.4$ to $t=40$
showing the way the pulse velocity deforms to produce the first
multipulse structure. Plots were shifted by 10 units for display
reasons.}} \label{panel3}
\end{figure}
%********|*********|*********|*********|*********|*********|*********|****

The scattering process is as follows: the solitary wave comes to
the interface and no backscattered solitary wave is observed but a
multipulse structure is generated as shown in Figure 6. This is a
remarkable phenomenon whose origin is not completely clear to us
(apart from the fact that it originates in the discreteness and
nonlinearity of the system in question). It was first observed by
Nesterenko and coworkers in \cite{Nesterenko2}.

The new effect that we report here is that a time after it appears
a second multipulse structure, with similar characteristics than
the first one but with less energy.  This is shown in Figure 7.

%********|*********|*********|*********|*********|*********|*********|****
%********|*********|*********|*********|*********|*********|*********|****
%********|*********| Fig6both:           |*********|*********|****
%********|*********|*********|*********|*********|*********|*********|****
\begin{figure}[h]
%\centering
\vspace{-.5cm} \hspace{-.5 cm}
\includegraphics[width=.35\textwidth]{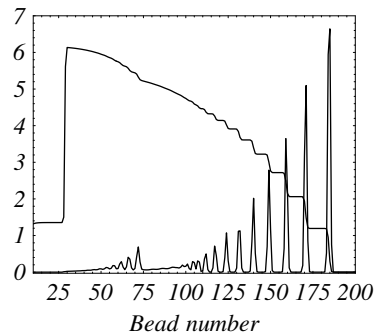} \vspace{-.2cm}
\caption{{Grain displacements with respect to their original
equilibrium position and pulse velocity as a function of bead
number at $t=115$. The pulse velocity has been scaled by a factor
6.}} \label{panel10}
\end{figure}
%********|*********|*********|*********|*********|*********|*********|****

Notice that there are some similarities between this effect and
the one observed before. Indeed, when the perturbation travels
from the light to the heavy system, the multipulse structures are
generated at the interface, leaving some energy behind the
interface. If we zoom-in Figure 7 we notice that something similar
happens here. Indeed, in Figure 8 and 9 we show how the secondary
multipulse structure is generated by a "seed" of energy that
remains at the interface after the primary multipulse structure
has been generated.

%********|*********|*********|*********|*********|*********|*********|****
%********|*********| Fig7inset:           |*********|*********|****
%********|*********|*********|*********|*********|*********|*********|****
\begin{figure}[h]
%\centering
\vspace{-.5cm} \hspace{-.5 cm}
\includegraphics[width=.35\textwidth]{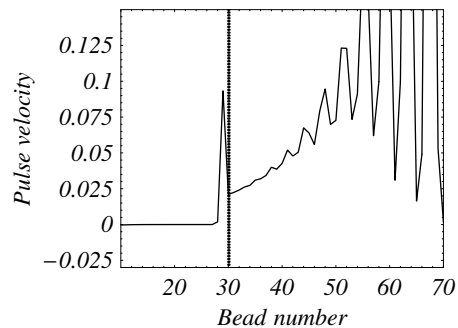}
\vspace{-.2cm} \caption{{Zoom-in of Fig. 7, at time $t=60$,
showing the "seed" of energy behind the behind the interface and
part of the primary multipulse structure.}} \label{panel5}
\end{figure}

%********|*********|*********|*********|*********|*********|*********|****
%********|*********| Fig7inset:           |*********|*********|****
%********|*********|*********|*********|*********|*********|*********|****
\begin{figure}[h]
%\centering
\vspace{-.5cm} \hspace{-.5 cm}
\includegraphics[width=.35\textwidth]{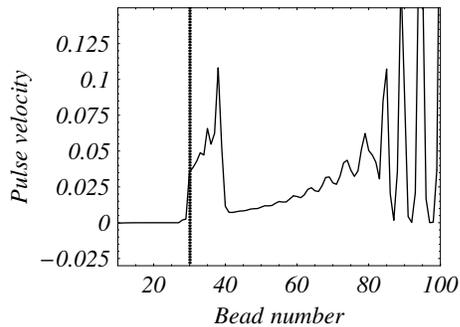}
\vspace{-.2cm} \caption{{Zoom-in of Fig. 7, at time $t=88$,
showing the generation of the secondary multipulse structure.}}
\label{panel5}
\end{figure}

Using a detailed numerical approach, we have studied the
scattering of solitary waves in a system with to "sonic vacua" in
a granular media with Hertzian contact. We have studied the same
system as the one analyzed in \cite{Nesterenko2} (except for the
existence of a wall, that makes no difference in the final
results) and found that, in addition to what observed there,
multipulse structures with smaller energies emerge. As far as we
know, this kind of secondary multipulse structures have not been
observed, yet, in experiments. We are aware of a new experimental
method suitable for the investigation of solitary waves at walls
(we thank Prof. Francisco Melo for giving us a copy of his paper
\cite{Melo} after the completion of this work). We expect that
this method could be used for studying the effects shown here.

Although we have given some information about the formation of the
secondary multipulses structures shown here, their origin is not
completely clear to us but it is interesting to notice that
similar multipulse structures appear in the scattering of two
solitary waves \cite{SenMan2} (although in that case only primary
multipulse structures appear). We agree with them that part of the
explanation lies in the discreteness of the medium through which
the solitary waves propagate.

Finally, it is worth to mention that although the effects shown
here are small, in our opinion they are important in the sense
that they form part of the dynamics of the scattering of solitary
waves by interfaces and then deserve to be studied and their
origin further clarified.

%%%%%%%%%%%%%%%%%%%%%%%%%%%%%%%%%%%%%%%%%%%%%%%%%%%%%%%%%%%%%%%%%%

\subsection*{Acknowledgements}

I thank Dr. St\'ephane Job for a talk that originated my attention
on this interesting subject and his interest in preliminary
results that encouraged me to go forward. I want to acknowledge
fruitful discussions with Dr. Ra\'ul Labb\'e and Dr. St\'ephane
Job. I would specially like to thank useful and encouraging
correspondence with Prof. Vitali F. Nesterenko.
%%%%%%%%%%%%%%%%%%%%%%%%%%%%%%%%%%%%%%%%%%%%%%%%%%%%%%%%%%%%%%%%%%%%%

\end{document}